\documentclass[]{elsart}

\usepackage{natbib}

\usepackage{graphicx}

\newcommand{\BibTeX}{ \textrm{B\kern-.05em\textsc{i\kern-.025em b}\kern-.08em
    T\kern-.1667em\lower.7ex\hbox{E}\kern-.125emX} }

\begin{document}

\begin{frontmatter}



\title{The spatial distribution of polar hydrogen deposits on the Moon}


\author[vre]{V. R. Eke}, 
\author[lfat]{L. F. A. Teodoro},
\author[rce]{R. C. Elphic}

\address[vre]{Institute for Computational Cosmology, Physics Department, Durham University, Science Laboratories, South Road, Durham DH1 3LE (U.K.)}
\address[lfat]{Astronomy and Astrophysics Group, Department of Physics and Astronomy, Kelvin Building, University of Glasgow, Glasgow G12 8QQ (U. K.)}
\address[rce]{Planetary Systems Branch, Space Science and Astrobiology
  Division, MS: 245-3, NASA Ames Research Center, Moffett Field, CA 94035-1000 (U.S.A.)}



%
%
%
%
%


\end{frontmatter}




\section*{Abstract}

A new analysis of the Lunar Prospector epithermal neutron data is
presented, providing an improved map of the distribution of hydrogen
near to the lunar poles. This is achieved using a specially developed
pixon image reconstruction algorithm to deconvolve the
instrumental response of the Lunar Prospector's neutron
spectrometer from the observed data, while simultaneously suppressing
the statistical noise. The results show that these data alone require the
hydrogen to be concentrated into the cold traps at up to 1 wt\%
water-equivalent hydrogen. This combination of localisation and high
concentration suggests that the hydrogen is present either in the form
of a volatile compound or as solar wind protons implanted into small
regolith grains.

\section{Introduction}

Is there ice on the Moon? Finding the answer is important
both for our understanding of the Solar System and for planning 
future visits.
The fact that the Moon's axis of rotation is inclined at only $1.6^\circ$
from the normal to the ecliptic means that permanently shaded regions
exist inside many 
craters within $10^\circ$ of the lunar poles. These regions
should be sufficiently cold ($<100$K) to prevent significant
sublimation of water ice over billions of years \citep{vas99}. The flux
of water reaching the Moon via 
impacts with comets and meteorites and subsequently able to
migrate to the relative safety of the polar `cold
traps' should be significant over the age of the solar
system \citep{arn79,cv03,but97}. Hydrogen delivered by 
the solar wind can also create water by chemical reduction of the
lunar regolith, providing another potential source
of polar ice \citep{arn79,cv00}.

While the existence of ice on the Moon was considered in the
1960s \citep{wat61}, interest was rekindled in the 1990s, when two
different methods were employed to address the issue, namely radar reflection
experiments and neutron spectroscopy from Lunar Prospector. 
The radar
method relies upon detecting the preferential reflection of the same sense of
circular polarisation carried by the incident radiation. This coherent
backscatter opposition effect could result from volume
scattering within a largely transparent medium, such as reasonably
pure ice \citep{hapke}. However,
lunar observations have shown that same-sense circularly polarised
radio waves are preferentially returned even from sunlit
locations \citep{stac97,st99,cam05,cam06}, making the most plausible explanation
that this is surface scattering from blocky ejecta within craters,
rather than volume scattering from large chunks of ice.

The strongest evidence for water ice at the poles is provided by the
deficit in the epithermal (0.3eV-100keV) neutron counts detected by the Lunar
Prospector. Cosmic rays impacting upon nuclei in the regolith liberate
neutrons. These elastically and inelastically scatter off nuclei in
the top few metres of the lunar 
regolith and some leak out of the surface before they suffer nuclear
capture or spontaneously decay.
Adding hydrogen to the regolith means that kinetic energy is more
rapidly drained from the neutrons during elastic
collisions. Consequently, fewer epithermal neutrons escape to be 
detected by orbiting neutron spectrometers. Thus, a deficit of 
epithermal neutrons is produced when there is a concentration of
hydrogen in the top $\sim 1$ metre of the regolith \citep{feld0}. 

Lunar Prospector's neutron spectrometer detected a
significant dip in the epithermal neutron count rate over both lunar
poles, suggesting an excess of hydrogen in these
areas \citep{feld1,feld2,feld3}. These tantalising results did not
definitively determine whether or not the hydrogen was confined to the
cold traps. If the excess hydrogen were distributed throughout the
polar regions, then it could just be solar wind-implanted protons
bound in OH groups that were less likely to diffuse out of the locally
cold regolith \citep{star00}. Even if the hydrogen were confined to
the cold traps, then the temperatures should be low enough that
regolith grains will retain some free H atoms. \cite{feld3} studied this
in detail, concluding that the diffusivity of hydrogen was unlikely to
be relevant within cold traps. When trying to discriminate between these various
possible forms for the excess polar hydrogen, it would clearly be
helpful to know whether or not it is confined to the permanently
shaded regions, and how concentrated this hydrogen is.

\section{Pixon Image Reconstruction}

Earlier studies of the Lunar Prospector data have used either the
high-altitude (100 km) epithermal data \citep{feld1} or smoothed
versions of the low-altitude (30 km) data \citep{feld2,feld3}. The
response of the Lunar Prospector has a FWHM of 130 km and 45 km for high and
low altitude orbits respectively. To determine whether or not the
hydrogen is confined to the cold traps requires an approach that 
deconvolves the Lunar Prospector data while not introducing spurious complexity.
This is the usual challenge in image reconstruction. Namely, given an
input two-dimensional map, $I$, which is measured with an instrument
that blurs it through a response function, $B$, and introduces
some random noise, $N$, then the measured data satisfies
\begin{equation}
D=I*B+N.
\end{equation}
The inclusion of noise means that the best
that can be achieved is that the inferred image, ${\hat I}$, avoids
unnecessary complexity and produces a residual field,
\begin{equation}
R=D-{\hat I}*B,
\end{equation}
that is statistically equivalent to the anticipated noise, $N$. 

The standard way to approach this problem is to consider the following
equation concerning conditional probabilities:
\begin{equation}
p({\hat I},M|D)=\frac{p(D|{\hat I},M)p({\hat I}|M)p(M)}{p(D)}.
\end{equation}
$M$ represents all aspects of the model for transforming the inferred
image, ${\hat I}$, to the data, $D$. This includes the assumed
response function, $B$, the pixellisation of the data and the assumed
characteristics of the noise. $p(D)$ is a constant, and, to avoid
introducing prejudice, so is $p(M)$. Thus,
\begin{equation}
p({\hat I},M|D) \propto p(D|{\hat I},M)p({\hat I}|M).
\end{equation}
The term on the left-hand side of this proportionality is commonly
known as the posteriori probability, 
$P_{\rm post} \equiv p({\hat I},M|D)$, which should 
be maximised by the reconstruction. The two terms on the right are the
likelihood ($p(D|{\hat I},M) \equiv e^{-0.5\chi^2}$ for the usual
definition of the $\chi^2$ goodness-of-fit statistic) and the image
prior, $P_{IP} \equiv p({\hat I}|M)$, respectively. The likelihood 
increases as the complexity of the reconstructed image, ${\hat I}$, is increased
and the fit to the data improves. In contrast, the image prior decreases with
increasing complexity in ${\hat I}$. This means that the optimal reconstruction
(maximum $P_{\rm post}$) occurs at the point where increasing
complexity in the reconstruction no longer improves the fit
sufficiently to be justified by the information content in the data.

A particularly suitable method for maximising the posteriori
probability is the pixon image reconstruction 
algorithm \citep{pp93,eke01}. A pixon can be thought of as a collection
of pixels carrying a particular amount of information. Thus, the
reconstruction of the Lunar Prospector data, which is well-sampled at
the lunar poles and less so further away due to the polar orbit of the
spacecraft, will have many small pixons to describe ${\hat I}$ near to
the pole, and a few larger pixons where the signal-to-noise ratio
decreases at lower latitudes. ${\hat I}$ is
constructed by convolving a pseudoimage, defined in the pixel array
and constrained to lie within physically acceptable count rates,
with a position-dependent kernel, whose width represents the local
pixon size. The optimum reconstruction is the one that combines
simplicity of the reconstructed image with an acceptable residual
field.

\subsection{Choosing the optimum number of pixons}
In order to determine the optimum number of pixons, $n_{\rm p}$, required to
reconstruct an image, the total information in the data set is first
determined via 
\begin{equation}
\Upsilon=\sum_{\bf x}
\frac{n_{\rm n}({\bf x})}{\alpha\sqrt{n_{\rm n}({\bf
      x})}}=\sum_{\bf x} \upsilon({\bf x}), 
\end{equation}
where $n_{\rm n}({\bf x})$ is the number of neutrons in pixel
{\bf x}, and $\upsilon({\bf x})=\sqrt{n_{\rm n}({\bf x})}/\alpha$ is the
signal-to-noise ratio in
pixel {\bf x}. Note that the choice of $\alpha$ is somewhat arbitrary. How
many $\sigma$ should a signal be before it counts as a piece of
information? A value of $\alpha=2$ has been chosen, although no
significant change is made to the reconstruction by choosing $\alpha=1$.

To maximise the entropy of the 
reconstruction, the information placed into each pixon should be the same.
Thus, for a given total number of pixons
in the reconstruction, each should contain $\Upsilon/n_{\rm p}$
pieces of information. The  Gaussian pixon, $K_{\bf x}$, was selected
for each pixel such that
\begin{equation}
\upsilon*K_{\bf x}=\Upsilon/n_{\rm p}.
\end{equation}
With the pixon distribution set for a particular total number of
pixons, the algorithm then uses the Fletcher-Reeves conjugate gradient
algorithm to minimise $E_r$, the autocorrelation of the reduced
residuals ($r=R/\sigma$) statistic \citep{pp92}, applied for lags of 1 pixel. 
Having thus found the best fit
for this number of pixons, the posteriori probability, $P_{\rm post}$,
can be calculated by multiplying together the likelihood
($e^{-0.5\chi^2}$) and the image prior, which is
\begin{equation}
P_{IP}=\frac{\Upsilon!}{n_{\rm p}^\Upsilon 
\left[\left(\frac{\Upsilon}{n_{\rm p}}\right)!\right]^{n_{\rm p}}}.
\end{equation}
Using Stirling's approximation gives
\begin{equation}
{\rm ln}(P_{\rm post})\approx 0.5[{\rm ln}\Upsilon+(1-n_{\rm p}){\rm ln}(2\pi)
-n_{\rm p}{\rm ln}(\Upsilon/n_{\rm p})-\chi^2].
\end{equation}
Thus, the number of pixons that maximises ${\rm ln}(P_{\rm post})$ can
be found, and this provides the optimal reconstruction.

\subsection{Coupled and decoupled reconstructions}
The reconstructed image, $\hat I$, is defined by gathering signal from
a pseudoimage, $W$. This gathering amounts to a convolution of the
pseudoimage with the normalised pixon kernel appropriate for each
pixel, namely 
\begin{equation}
\hat I ({\bf x}) = W*K_{\bf x}.
\end{equation}
Constraints are placed on $W$, so that it cannot take unphysical
values, such as count rates less than zero or greater than 22
epithermal neutrons per second, which is larger than the rate expected
from regolith containing no hydrogen whatsoever \citep{law06}.
A set of 11 Gaussian pixons, with widths varying from $10$ km to
$250$ km are available for the algorithm to choose. When a single
pseudoimage is employed, the sunlit and permanently shaded pixels are
treated equally. This is termed a coupled reconstruction. However, one
should include some
prior knowledge that sunlit pixels would not be expected to have a
hydrogen content exceeding a few times that in the returned lunar
samples. A loose limit of $15$ counts per second is chosen for
the minimum allowable epithermal rate in sunlit pixels. Introducing
an independently determined map of permanently shaded
pixels enables a decoupled treatment of the two types of
pixel; sunlit and shadow. Each pixel type has its own pseudoimage,
with different limits on their allowed count rates: sunlit - [15,22]
and shadow - [0,22]. These pseudoimages only 
have non-zero elements for pixels of the type being considered. Thus,
to avoid having the pseudoimage grow to compensate for this near to
the boundaries between pixel types, a mask, $m_{\rm type}({\bf x})$, of ones and
zeros is set up for each pixel type, with the ones lying only in
pixels of the chosen type. This mask is then convolved with 
the appropriate pixon kernel for each pixel to give a map of weights
\begin{equation}
M_{\rm type} = m_{\rm type}*K_{\bf x}.
\end{equation}
The resulting map, $M_{\rm type}$, has values lying in the range $(0,1]$ and can
be used to define the decoupled reconstruction via
\begin{equation}
\hat I = \frac{(W_{\rm type}*K_{\bf x})}{M_{\rm type}}.
\end{equation}
In practice because the required pixon size is a continuous variable,
whereas only a finite set of distinct pixon sizes are actually used, the
reconstruction is created by interpolating between the nearest two
available pixon kernels. By decoupling the shadow and sunlit pixel
values, the effective number of pixons being used increases. To
account for this, $n_{\rm p}$ is replaced in the image prior
calculation by $n_{\rm p}~G$, where
\begin{equation}
G=\frac{1}{n_{\rm pixel}}\sum_{\bf x}\frac{1}{M_{\rm type}({\bf x})}~,
\end{equation}
with the sum being over all $n_{\rm pixel}$ pixels ${\bf x}$.

Note that, for the decoupled reconstructions
of the real data, the sunlit pixels take values in the ranges
[18.5,20.5] counts per second for the north pole, and [18.3,20.7]
counts per second for the south pole. For the north pole, this range
corresponds to plausible amounts of hydrogen for occasionally sunlit
terrain. The construction of the north and south pole shadow maps is
described in detail by \cite{rick07}. In brief, radar tomographic maps
(Margot et al., 1999; U.S. Geological Survey, 2002) were combined with
the model of polar craters produced 
by \cite{buss03}. Gaps were filled by interpolation and the whole
resulting landscape was projected on a sphere of 1738 km radius, and
illuminated with summer sunlight. Note that the permanently
shaded areas are better determined around the north pole because the south
pole shadow map was largely derived from
data taken during winter lighting conditions \citep{rick07}. The less
well determined shadow map in the south may be 
behind the larger range of values being used by the sunlit pixels in
this case. 

\begin{figure}
\begin{center}
\hspace{-0.5cm}\includegraphics[width=5.5in]{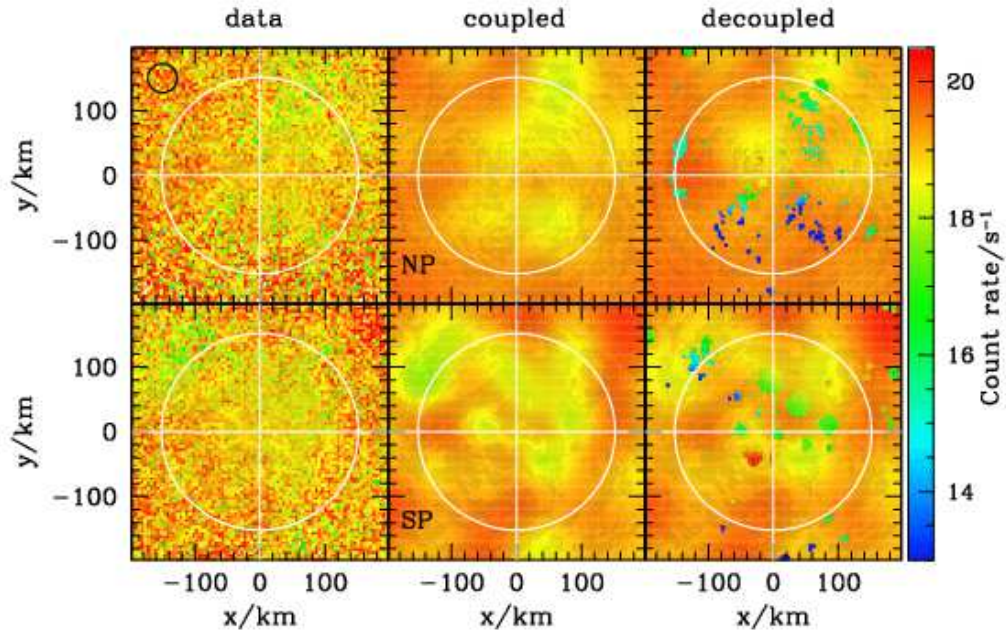}
\caption[Epithermal count rate maps]{
	\label{epimaps}
	Epithermal count rate maps for the north (top row) and
        south (bottom row) poles. From left to right, the columns represent
        the data, $D$, the coupled and decoupled reconstructions,
        $\hat{I}$. The white circles represent a latitude of $\pm 85^\circ$
        and the shading shows Clementine imaging. Also shown, with a
        black circle in the top left panel is the size of the Lunar
        Prospector's PSF. The spherical surface has been represented in
        these panels such that the distance from the image centre
        represents the arc length to the pole, and, following
        convention, $0^\circ$ longitude points down/up for the
        north/south poles respectively.}
\end{center}
\end{figure}

The left hand panels of Fig\ref{epimaps} show the north (top) and south
(bottom) pole epithermal data. These can be contrasted with the
central and right hand columns in Fig\ref{epimaps}, showing ${\hat I}$ 
from two different reconstructions of both data sets.
These four reconstructions all provide residual fields with
acceptable $\chi^2$ values, and differ only because an additional shadow
map prior was included for the decoupled reconstructions in the right hand
column, unlike the coupled reconstructions in the central column. 
The coupled reconstructions are very similar to the count rate
maps created by merely smoothing the observational data using a
Gaussian with FWHM of 40 km, as chosen by \cite{feld3}. The decoupling
significantly changes the inferred hydrogen abundance within the
cold traps, which are clearly visible because of the rapid count rate
jumps across their edges. If a good fit is defined merely using
$\chi^2$, then there 
appears to be no need to impose the shadow map. The coupled
reconstruction is simpler too, so the use of the shadow map can only
be justified if the data demand it.

\section{Results}

The coupled and decoupled reconstructions differ significantly only in
the epithermal neutron count rates present in the cold traps. Thus,
the residuals in the vicinity of these permanently shaded areas offer
the best route to determine if the data contain sufficient information
to discriminate between these
two different reconstructions. If a real count rate dip in a cold trap
were wrongly smoothed over in the reconstruction, then one would
expect to see a radial variation in the mean residual moving out from the
cold trap centre. The pattern of these spatially correlated residuals
would depend upon the size of the cold trap, the depth of the count
rate dip, and the FWHM of the instrumental response.

\subsection{The radial dependence of the mean stacked residuals}

In order to try
and discriminate between the two types of reconstruction, it is
desirable to stack together the residuals around all sufficiently
large (at least 3 contiguous pixels) cold traps. From the right
hand panels in Fig\ref{epimaps}, it is clear that the cold traps
have a variety of sizes and shapes, and that they are often
sufficiently close to one another that the anticipated rings of
non-zero residuals around them would overlap. In the face of these
complications, the easiest 
way to understand the significance of the results is to use a Monte
Carlo method, whereby mock data sets are created and then analysed in
exactly the same way as the real data itself. To this end, 100
different noisy mock Lunar Prospector time series data sets were
created, taking the reconstructions in Fig\ref{epimaps} as the input 
images, $I(\bf x)$. The
different mocks varied only as a result of the random numbers
chosen to add the noise. Both coupled and decoupled
reconstructions were performed, and the stacked radial profiles of the reduced
residuals, $r=R/\sigma$ with $\sigma$ coming from counting statistics, 
around cold trap centres were calculated. Pixels positioned near to more
than one permanently shaded region were only counted once, at the
radius appropriate for the nearest cold trap.

\begin{figure}
\begin{center}
\includegraphics[width=5.5in]{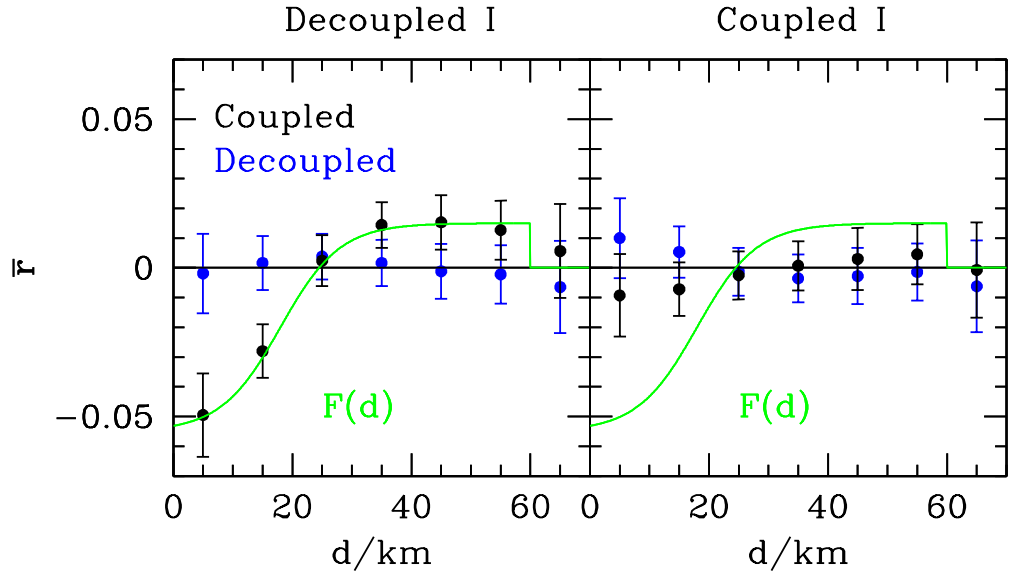}
\caption[Mock radial variation of residuals around cold traps]{
	\label{rings}
	The variation of the mean reduced residual per
  observation, ${\bar r}$, with the distance to the centre of the nearest 
  permanently shaded crater, $d$. The results are stacked for all permanently 
  shaded areas in both the north and south polar regions containing at
  least 3 contiguous shadow pixels. Coupled (black) and
  decoupled (blue) reconstructions are shown in both panels.
  The left hand panel shows reconstructions from mock data made with
  the decoupled reconstruction of the real data as the input
  image. Error bars represent the $1\sigma$ scatter between the individual 
  mock results. The green curve represents the function $F(d)$ given in
  Eq. \ref{fdef}, which is a fit to the results from the coupled
  reconstructions. 
  The right hand panel shows the results for mocks made using the
  coupled reconstructions of the real data, shown in the central
  column of Fig\ref{epimaps}, as the input image for the 100 mock catalogues.}
\end{center}
\end{figure}

The left hand panel in Fig\ref{rings} shows the radial
dependence of the reduced residuals for both types of reconstruction
of the mocks 
created from the decoupled reconstructions of the real data. Error
bars represent the $1\sigma$ scatter between the results obtained from the
individual mock data sets. The decoupled reconstructions show 
no radial trend in the mean reduced residual around the cold traps,
which contrasts with the results from the coupled
reconstructions. These show a negative residual for distances less
than $\sim 25$ km, swapping over to a positive one at larger radii,
signifying that ${\hat I}$ is too large in the cold traps. While the
global $\chi^2$ may be acceptable, the coupled
reconstruction produces spatially correlated residuals near to cold traps.
To remove these would require either a vastly increased
number of pixons, which would produce a hugely detailed and improbable
reconstruction, or the imposition of a shadow map. In short, these mock
Lunar Prospector epithermal neutron data sets contain sufficient
information to distinguish between the coupled and decoupled
reconstructions, i.e. the real data should tell us if the excess
hydrogen is concentrated into the cold traps or more diffusely distributed.

The right hand panel in Fig\ref{rings} shows how the mean reduced
residuals vary when the pixon algorithm is applied to mock data
sets constructed from the input image where the count rate dips are
not localised to the crater, i.e. the coupled reconstruction of the
real data. It
is apparent that there is a small bias, whereby the decoupled
reconstruction still tries to put a slight count rate dip into the cold
traps where it is not actually present. Also, the coupled reconstruction
tends to
oversmooth the small dips that did exist. However, these systematic
biases resulting from the method are
negligible relative to the size of the deviations found in the case of
the decoupled input image.

\begin{figure}
\begin{center}
\includegraphics[width=5.5in]{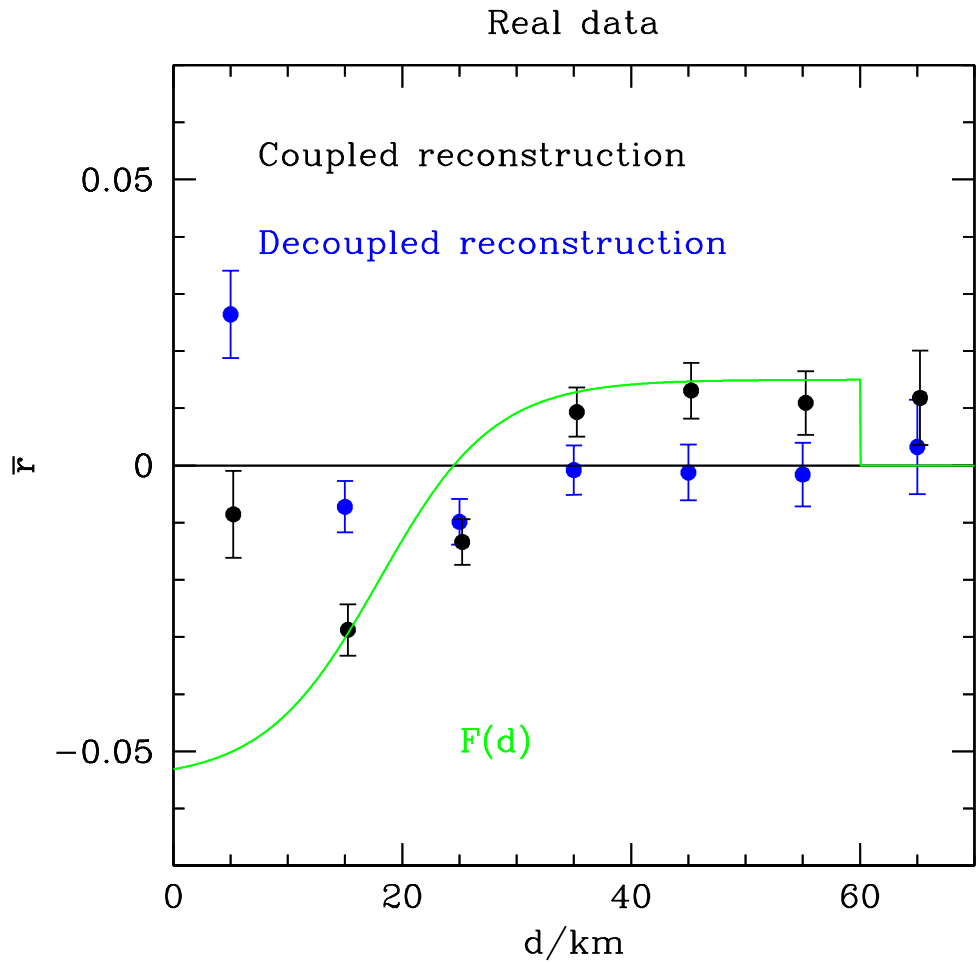}
\caption{As for Fig\ref{rings}, using the stacked reduced
        residuals around cold traps for reconstructions of the real
        data. Error bars represent the errors on the mean reduced residuals.}
\label{ringc}
\end{center}
\end{figure}

Fig\ref{ringc} shows the corresponding information for the coupled
and decoupled reconstructions of the real data, with the error bars
now representing the error on the mean reduced residual.
A very similar pattern in the residuals is
seen, for all but the central bin, as was the case for the mock
reconstructions made using the decoupled input image. There are
relatively few pixels, and hence 
observations, contributing to the central bin, which nevertheless
still lies in the tail of the distribution of the mock results.
Just as for the mock reconstructions, including the independently
determined shadow map removes the correlated residuals out to $60$ km
from the cold trap centres. Thus, the Lunar Prospector data are 
not adequately described by the coupled reconstruction, and they do
require that the epithermal count rate dips are concentrated into the
permanently shaded regions.

\subsection{Shadowiness}
\begin{figure}
\begin{center}
\hspace{-0.5cm}\includegraphics[width=5.5in]{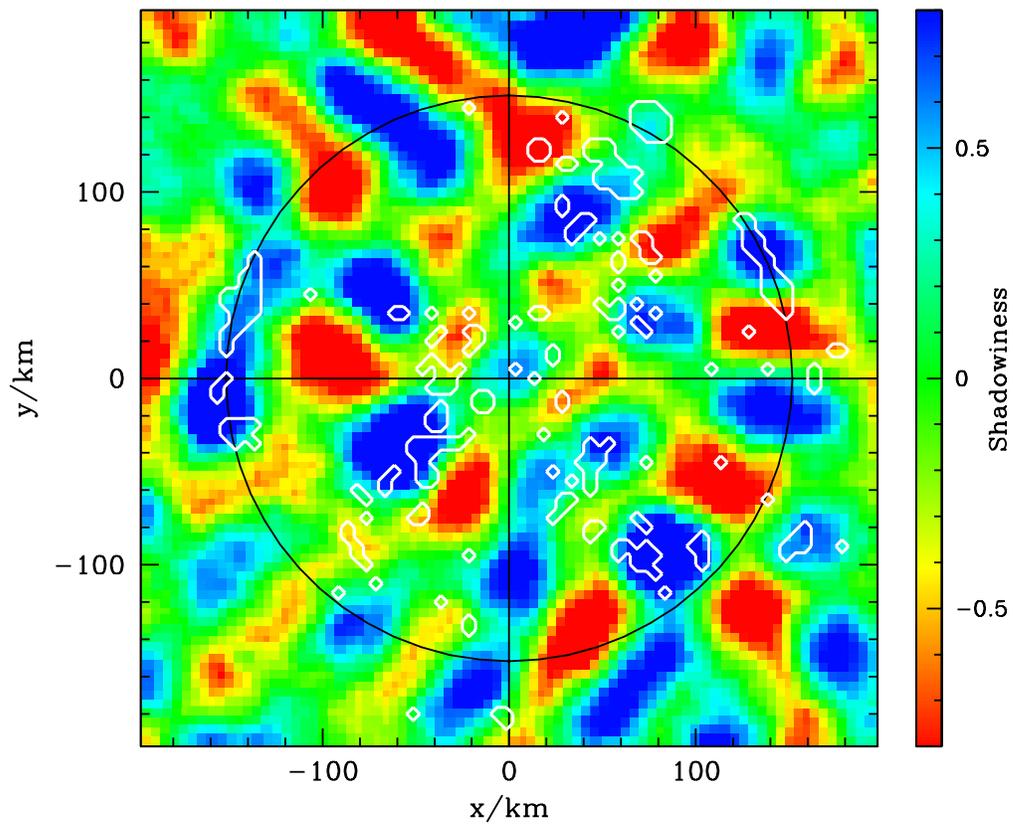}
\caption[North pole shadowiness map]{
	\label{smap}
        The shadowiness map for the coupled
        reconstruction of the north pole Lunar Prospector data. Cold traps
        are outlined in white, and these can be seen to lie preferentially
        in regions of high shadowiness according to the neutron data.}
\end{center}
\end{figure}

The results in the previous subsection imply that the Lunar Prospector
epithermal neutron data themselves pick out 
the pixels that should be cold traps. In this section, the same
conclusion is reached in a more quantitative way without resorting to
stacking together the disparate cold traps. Consider
a shadowiness parameter $S$ defined via
\begin{equation}
S=r*F,
\end{equation}
where the filter function $F$, shown by the green curve in Fig\ref{rings},
is given by
\begin{equation}
\label{fdef}
F(d)=\left\{ \begin{array}{lll}
  0.035~{\rm tanh}\left( \frac{(d/{\rm km})-18}{10} \right)-0.02 & ~~~
  & d\le 60~{\rm km};\\
  0 & & d>60~{\rm km}. \\
  \end{array}
\right.
\end{equation}
$F(d)$ is merely a function that represents the radial dependence of the
spatial correlations in 
the reduced residuals when a coupled reconstruction is performed
on a data set derived from an input image where the hydrogen is
concentrated into shaded areas. For spatially uncorrelated residuals,
the distribution of $S$ should have a median of zero. A significantly
positive median value would imply that the reconstruction is
systematically overestimating the count rate in that set of pixels.
A map of the shadowiness parameter is shown in Fig\ref{smap} for the coupled
reconstruction of the real north pole data, where the shadow map prior
is relatively well known. The locations of the independently
determined shadow pixels are 
superimposed in white. Areas of high shadowiness, shown in blue,
tend to coincide with the permanently shaded regions.

\begin{figure}
\begin{center}
\hspace{-0.5cm}\includegraphics[width=5.5in]{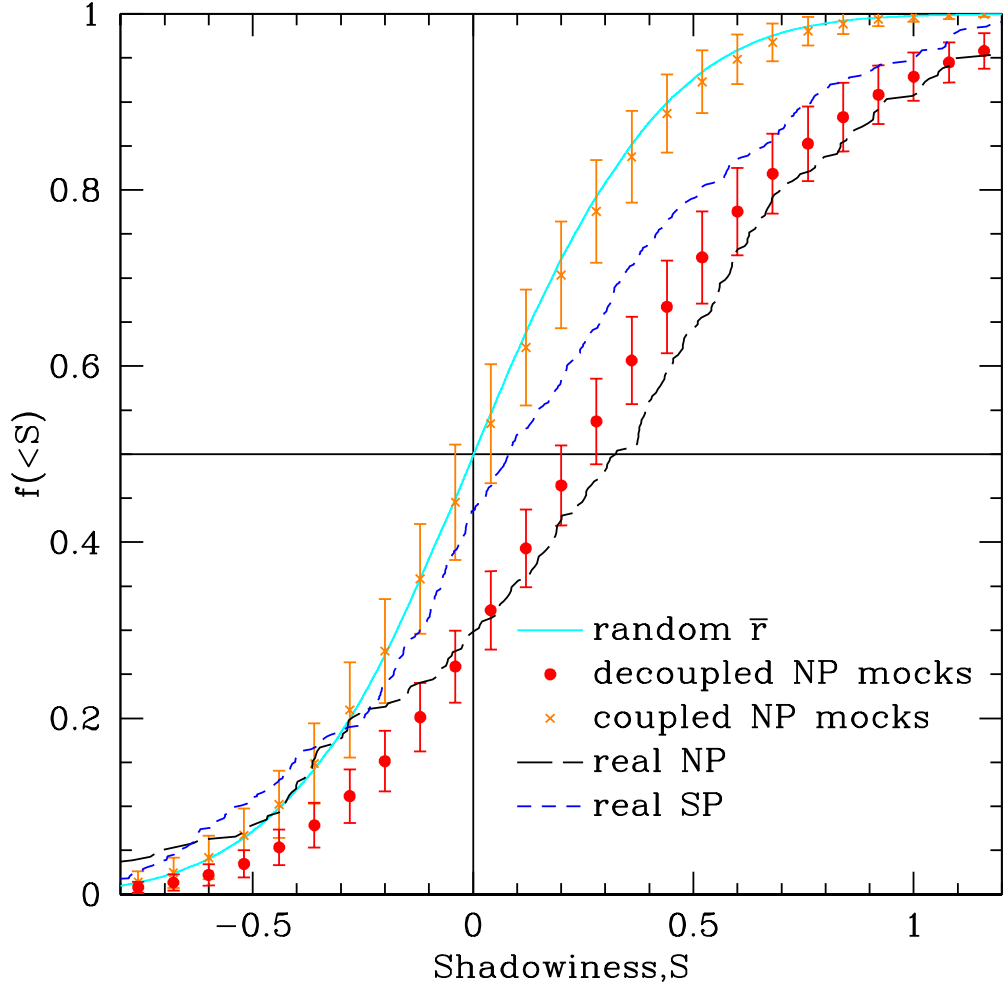}
\caption[Pixel shadowiness distributions]{
	\label{sdist}
	\label{lastfig}			
        Shadowiness distributions for various coupled
  north pole reconstructions. Points show mean results from 100 mock
  data sets made using either the coupled (crosses) or decoupled
  (circles) input image. Error bars represent the standard deviation
  between the mock distributions. The distributions of the mock results
  are accurately represented by Gaussians. Results from the north and south pole
  reconstructions of the real data are shown by long- and short-dashed
  lines respectively. A solid line shows what the shadowiness
  distribution looks like for an uncorrelated random Gaussian residual
  field with $\sigma=1$.}
\end{center}
\end{figure}

Fig\ref{sdist} shows the cumulative distributions of shadowiness for a
random reduced residual field, and for the shadow pixels for a number
of different coupled reconstructions. The points show the mean distribution
recovered from 100 mocks made using either the decoupled (circles) or
coupled (crosses) input north pole image. Error bars represent the standard
deviation between the mock reconstructions. The long-dashed line is
the distribution for the real north pole shadow pixels.
If the shadowiness parameter and the shadow map
were truly uncorrelated, then this cumulative distribution of $S$ would be
indistinguishable from that resulting from the randomly distributed
residuals. This is not the case, implying that the Lunar Prospector
data have picked out as special, in a statistical sense, the pixels
that are independently 
known to be permanently shaded. While not as convincing, perhaps as a result of
the less well determined shadow map, the south pole results have also
been added with a short-dashed line. 

\subsection{Uncertainties in the reconstructions}

While the stacked reduced residuals from all north and south pole
shaded regions contain enough information to determine whether or not
the count rate dips are typically concentrated into the cold traps, it is also
interesting to know what is the uncertainty on the mean count rate
within any individual cold trap. Permanently shaded regions 
that received more observations,
either as a result of being large or near to the poles, will have
better constrained mean epithermal neutron count rates. One would
really like to know, given the measured cold trap count rates, what is the
probability distribution of the true count rates. The mock
catalogues allow two slightly different, yet closely related,
questions to be answered, namely:\\
1. if the hydrogen is as concentrated as the decoupled reconstruction
of the real data suggests, then what are the bias and scatter in the
decoupled reconstruction values, and\\
2. if the hydrogen is as diffuse as the coupled reconstruction
of the real data suggests, then what is the probability that a
decoupled reconstruction will yield a cold trap count rate at least as
low as did that from the decoupled reconstruction of the real data.\\
Table \ref{tab1} contains this information, calculated from the
appropriate set of 100 mocks for a few notable craters. From these
data, it appears as if the north pole contains permanently shaded
areas with deeper dips, i.e. more convincing hydrogen excesses,
than exist in the south. An epithermal count rate of $\sim 13.5$ per second
corresponds to $1$ wt\% water-equivalent hydrogen \citep{law06}.

\subsection{Discussion}

The results presented in this section demonstrate that:\\
1. the Lunar Prospector epithermal neutron data rule out the hypothesis
that the hydrogen concentration is smoothly distributed in the polar
regions, and\\
2. the set of permanently shadowed pixels defined by the shadow maps
of \cite{rick07} is statistically picked out as special.\\
The first of these statements follows from the fact that spatially correlated
residuals result if one assumes that the hydrogen distribution is
smoothly distributed throughout the polar regions. These statistically
unacceptable residuals are evident in both Fig\ref{ringc} and
Fig\ref{sdist}. 
Over most of the reconstructed image, the residuals
are statistically acceptable and spatially uncorrelated. However,
these figures show the problem because they focus on the residuals in
the vicinity of a particular set of pixels, namely those that are
deemed to harbour regions of permanent shade. 
The information content latent within the data is sufficient to reject the
hypothesis that the hydrogen is smoothly distributed, but one needs to
know where to look to discover this message. The shadow map merely
acts to focus attention into the regions where the useful information resides.
If a random set of pixels
were chosen as centres around which to investigate the residuals, then
these correlations would not be seen. In fact, mapping the south pole
shadow map pixel locations onto the north pole data and looking at the
residuals around this different set of pixels shows no problem 
at all. The shadowiness distribution is completely
consistent with that expected if the residuals were really random and
spatially uncorrelated. Similarly, the residuals around a set of
pixels in north pole craters supposedly without areas of permanent
shade also show no deviations from the results that true randomly
distributed residuals would produce. These observations illustrate the
second statement above. The areas that are independently thought to
harbour permanent shade lie preferentially in regions where the
smoothly-distributed hydrogen reconstruction produces spatially
correlated residuals. The simplest, and only plausible, way to remove
these local correlations in the residuals is to concentrate the
hydrogen into the cold trap pixels.

\section{Conclusions}

This study shows that the Lunar Prospector data
alone require the polar hydrogen excess to be concentrated into the
permanently shaded cold traps, rather than being more diffusely
distributed. In some craters the concentration of hydrogen is
sufficiently high that it corresponds to $\sim 1$ wt\% water. This
combination of localisation into the permanently shaded polar craters
and consequent relatively high concentration is consistent with the
hypothesis that the hydrogen is present in the form of water ice. However,
these concentrations are not sufficiently high to preclude the
hydrogen being in the form of solar wind protons implanted into
regolith grains \citep{feld3}. While the Lunar Prospector data are not able to
determine the intracrater distribution of hydrogen, radar results
suggest that it is unlikely to be concentrated into small patches of high
purity water ice, at least not in the regions of
the cold traps accessible to Earth-based detectors. Assuming that the
excess hydrogen is in the form of water ice that reaches 2 m into the
regolith, the depth expected to be gardened in 2 billion years
\citep{feld3}, this
suggests that there are a few times $10^{11}$kg of water ice within
permanently shaded regions within $10^\circ$ of the lunar poles.

\appendix

\ack
We would like to thank Bill Feldman for helpful discussions.
V.R.E. acknowledges the support of a Royal Society University
Research Fellowship. L.F.A.T. acknowledges the support of a Leverhulme
Research Fellowship. R.C.E. acknowledges the support of a NASA Lunar
Reconnaissance Orbiter Participating Scientist Grant.

\label{lastpage}


\def\jgr{J. Geophys. Res. }
\def\grl{Geophys. Res. Lett. }
\def\nat{Nature }
\def\mnras{Mon. Not. R. Astron. Soc. }
\def\pasp{Publ. Astron. Soc. Pac. }

\bibliography{mybib}

\bibliographystyle{elsart-harv}


\clearpage	

\begin{table}
\begin{center}
\textbf{Mean epithermal neutron count
rates and their uncertainties for a few notable locations\\}
\vspace{0.5cm}
\begin{tabular}{l|c|c|c|c|c}\hline
Crater & Location & $d$ & $c$ & $d_{\rm mock}\pm 1\sigma$ & $P(\le d|c)$ \\ \hline
Cabaeus & $84.5^\circ {\rm S},322^\circ {\rm E}$ & $14.8$ & $18.1$ & $14.1\pm 1.1$ & $0.04$\\
de Gerlache & $88.5^\circ {\rm S},273^\circ {\rm E}$ & $16.8$ & $18.5$ & $17.5\pm 0.8$ & $0.22$\\
Faustini & $87.3^\circ {\rm S},77^\circ {\rm E}$ & $17.2$ & $18.4$ & $17.0\pm 0.7$ & $0.22$\\
Shackleton & $89.7^\circ {\rm S},110^\circ {\rm E}$ & $16.7$ & $18.8$ & $18.8\pm 0.5$ & $<0.01$\\
Shoemaker & $88.1^\circ {\rm S},45^\circ {\rm E}$ & $17.8$ & $18.7$ & $18.4\pm 0.4$ & $0.01$\\
Unnamed & $89.2^\circ {\rm N},36^\circ {\rm E}$ & $10.5$ & $18.6$ & $10.0\pm 1.3$ & $<0.01$\\
Unnamed & $89.6^\circ {\rm N},45^\circ {\rm E}$ & $13.8$ & $18.7$ & $12.8\pm 1.1$ & $0.01$\\
\hline
\end{tabular}
\caption[Relevant Files from Elsevier]	
	{
	\label{tab1}	
	\label{lasttable}		
The
decoupled, $d$, and coupled, $c$, reconstruction mean cold trap
epithermal neutron count rates
are listed, in counts per second, along with the mean and scatter of
the decoupled reconstructions of the decoupled mocks, $d_{\rm mock}$,
and the probability of measuring a value at least as low as $d$, when
doing a reconstruction of an image where the true value is $c$.
	}
\end{center}
\end{table}

\clearpage


\end{document}